\def\year{2022}\relax
\documentclass[letterpaper]{article} 
\usepackage{natbib}
\usepackage{aaai22}  
\usepackage{times}  
\usepackage{helvet} 
\usepackage{courier}  
\usepackage[hyphens]{url}  
\usepackage{graphicx} 
\def\UrlFont{\rm}  
\usepackage{graphicx}  
\usepackage{float}
\usepackage{amsfonts}
\usepackage{multirow}
\usepackage{amsmath}
\usepackage{xcolor}
\title{\huge{Hybrid neural networks for on-device directional hearing}}

\author{Anran Wang$^1$, Maruchi Kim$^1$, Hao Zhang$^2$, Shyamnath Gollakota$^1$\\
{\normalfont $^1$University of Washington, $^2$ETH Z\"urich}\\
{\normalfont \{anranw,mkimhj,gshyam\}@cs.washington.edu, hao.zhang@inf.ethz.ch}}

\begin{document}

    \maketitle
{\bf Abstract ---} On-device directional hearing requires audio source separation from a given direction while achieving stringent human-imperceptible latency requirements. While neural nets can achieve significantly better performance than traditional beamformers, all existing models  fall short of supporting low-latency causal inference on computationally-constrained wearables.
We present HybridBeam, a hybrid model that combines traditional beamformers with a custom  lightweight neural net. The former  reduces the computational burden of the latter and also improves its generalizability, while the latter is designed to further reduce the memory and computational overhead to enable real-time and low-latency operations. Our evaluation shows comparable performance to  state-of-the-art causal inference   models  on synthetic data while achieving a 5x reduction of model size, 4x reduction of computation per second, 5x reduction in processing time and generalizing better to  real hardware data.
Further, our real-time hybrid model runs in 8~ms on mobile CPUs  designed for low-power wearable devices  and achieves an end-to-end latency of 17.5~ms.   

\section{Introduction}
Directional hearing is the ability to amplify  speech from a specific direction while reducing sounds  coming from  other directions. This has multiple applications ranging from medical devices to augmented reality and wearable computing. Directional hearing aids can help  individuals with hearing impairments who have increased difficulty hearing in the presence of noise and interfering sounds~\cite{doclo2010acoustic,brayda2015spatially}. It can also be combined with augmented reality headsets to customize the sounds and noises from different directions, e.g.,  sensors like gaze trackers can enable a wearer to be in a noisy room and amplify the speech from a specific direction, simply by looking at the source.

For decades, the predominant approach to achieving this goal was to perform beamforming~\cite{krim1996two, brayda2015spatially, chhetri2018multichannel}. While these signal processing techniques can be  computationally light-weight,  they have a limited performance~\cite{souden2010study,kumatani2012microphone}. Recent work has shown that  neural networks achieve exceptional  source separation in comparison~\cite{luo2019conv,jenrungrot2020cone} but are computationally expensive and to date, cannot run on-device on wearable computing platforms.

Directional hearing applications however impose stringent  computational, real-time and low-latency requirements that are not met by any existing source separation networks. Specifically, compared to other audio applications like tele-conferencing where latencies on the order of 100~ms are  adequate, directional hearing requires real-time audio processing with much more stringent latency requirements. For example, medical hearing aid research shows that we need a latency less than 20~ms to be tolerable~\cite{stone1999tolerable}. This is also true for augmented reality applications so that any modified sound sources are correctly synced, both visually and audibly, in the brain~\cite{gupta2020acoustic}.

These stringent low-latency constraints are challenging to meet on wearable and medical devices.   Directional hearing requires not only processing the continuous audio input stream, but also generating a continuous output stream  within these real-time constraints. While powerful GPUs and specialized inference accelerators (e.g., TPU) can speed up the network run-time~\cite{wang2019benchmarking}, they are usually not available on a wearable device given their power, size and weight requirements.  In fact, even the CPU capabilities and memory bandwidth available on wearables can be significantly constrained even compared to smartphones. For example, the iPhone 12 CPU is more than 10 times faster than that used in Google glasses and Apple watch.  Offloading computation to other devices (e.g., smartphone) is not an option given the additional wireless roundtrip latency  on the order of tens of milliseconds.

In this paper, we show for the first time that real-time directional hearing using deep learning can be achieved on  computationally-constrained platforms. Instead of designing an end-to-end neural net to perform the task, we create a hybrid model that combines lightweight beamforming algorithms with neural networks. Our key insight is that the beamforming algorithms can provide  spatial hints to a neural net that can drastically reduce the network complexity and its computational cost while achieving similar source separation performance to state-of-the-art  causal neural networks that are computationally expensive. At a high level, while neural networks are a powerful tool to approximate functions, approximating beamforming functions like matrix inversion can increase the network complexity making them computationally expensive. However,  traditional beamformers can provide useful theoretically-derived spatial hints to the neural net, in a computationally inexpensive manner.  

We present the network architecture shown in Fig.~\ref{fig:pipeline} where we first input the signals from the multiple microphones into three different beamformers. The output of the beamformers along with the original signals from the multiple microphones is then fed into a causal neural net model that is optimized for memory overhead and inference time on mobile CPUs.  Specifically, we use complex tensors throughout our network to reduce half of our model size while achieving a comparable accuracy. Compared to real-valued networks, complex representation also restricts the degree of freedom of the parameters by enforcing correlation between the real and imaginary parts, which enhances the generalization capacity of the model  since phase encodes essential spatial information. We also
design a combination of dilated and strided complex convolution stacks to reduce memory footprint and memory copy per time step while keeping a large receptive field. Finally, we simplify the temporal convolutional network  to run  more efficiently.    

Compared to state-of-the-art causal models, we are able to achieve comparable separation performance with a 5x reduction of model size, 4x reduction of computation, and 5x reduction of processing time, enabling it to run in real-time on a CPU suitable for wearable devices with low latency. We also evaluate our model trained entirely on simulated data on real data recorded in conference rooms using a smart glasses prototype with a custom six-microphone array and a gaze tracker, which achieves real-time, gaze-controlled directional hearing with an end-to-end latency of 17.5~ms.

\section{Related Work}

\subsubsection{Beamformers based on statistical signal processing.}

Beamforming techniques are designed to combine multi-channel sensor signals to achieve directionality. 
Linear filters can statistically create constructive interference at the direction of interest and destructive interference elsewhere. Non-adaptive beamformers such as Barlett (delay-and-sum) beamformers and superdirective beamformers construct a constant linear filter applied to the signal~\cite{krim1996two}. Adaptive beamformers such as minimum-variant distortionless-response (MVDR) beamformers and linearly constrained minimum variance (LCMV) beamformers additionally utilize spatial information from the mixture signal to inform the filter construction~\cite{souden2010study}. While they can be computationally inexpensive,  their separation result is limited since they use the spatial cues but do not efficiently capture the acoustic cues.

\subsubsection{Blind source separation.} Another classical problem formulation is  blind source separation  where each channel receives a different unknown linear combination of a few independent sound sources~\cite{comon2010handbook}. Without spatial hints such as directions, the problem formulation  is often under-deterministic but a few spatial clustering methods such as independent component analysis~\cite{haykin2005cocktail} and Gaussian mixture model~\cite{higuchi2017online} can obtain a solution assuming a small number of sound sources. Recently, 
neural network architectures have been proposed to achieve blind source separation. Frequency-domain approaches  learn the frequency-time mask for each sound source that is applied to the mixture spectrogram~\cite{chen2018cracking,hu2020dccrn}. The spectrogram as input and output makes it inefficient for use in our application since it requires lookahead of tens of milliseconds. Time-domain approaches such as Demucs~\cite{facebookpaper}, TasNet~\cite{luo2018tasnet} FasNet~\cite{luo2019fasnet}, TAC~\cite{luo2020end}, Conv-TasNet~\cite{luo2019conv} and its variants\cite{gu2019end,defossez2019music,luo2020end,han2020real} optimize for the learnt filters that convolve with the mixture signals to separate each sound source. While  time-domain approaches allow causal construction and more effective separation, they are not designed to match directions with each separated signal from the mixture, and the computation grows exponentially with the number of sources. \cite{jenrungrot2020cone}  simultaneously separates each sound source as well as  identify their  directions; their  model however is not causally constructed. They are also an overkill for use with directional hearing since unlike blind source separation it only requires separating the speech from a specific direction.

{\subsubsection{Neural beamformers.} To address the specific problem where the direction is provided as input,  beamformer designs have been proposed using neural networks.   \cite{chen2018multi} presents a multi-pass bi-directional LSTM network using spectral, spatial, and angle features. Similarly, \cite{gu2019neural} design a LSTM network on the spectrogram and attention mechanisms. Neither of these networks are causal in structure. \cite{gu2020temporal} extends Conv-TasNet~\cite{luo2019conv} by feeding spatial features along with the first channel and achieves better results than LSTM-based methods,  {\cite{qian2018deep} uses a combination of beamformers and neural networks to enhance speech}. They are computationally expensive and does not meet the delay requirements of mobile CPUs. {\cite{fedorov2020tinylstms} achieves low-power speech enhancement using LSTM, but it is not for multichannel source separation.}

\begin{figure}[t]
    \centering
    \includegraphics[width=0.48\textwidth]{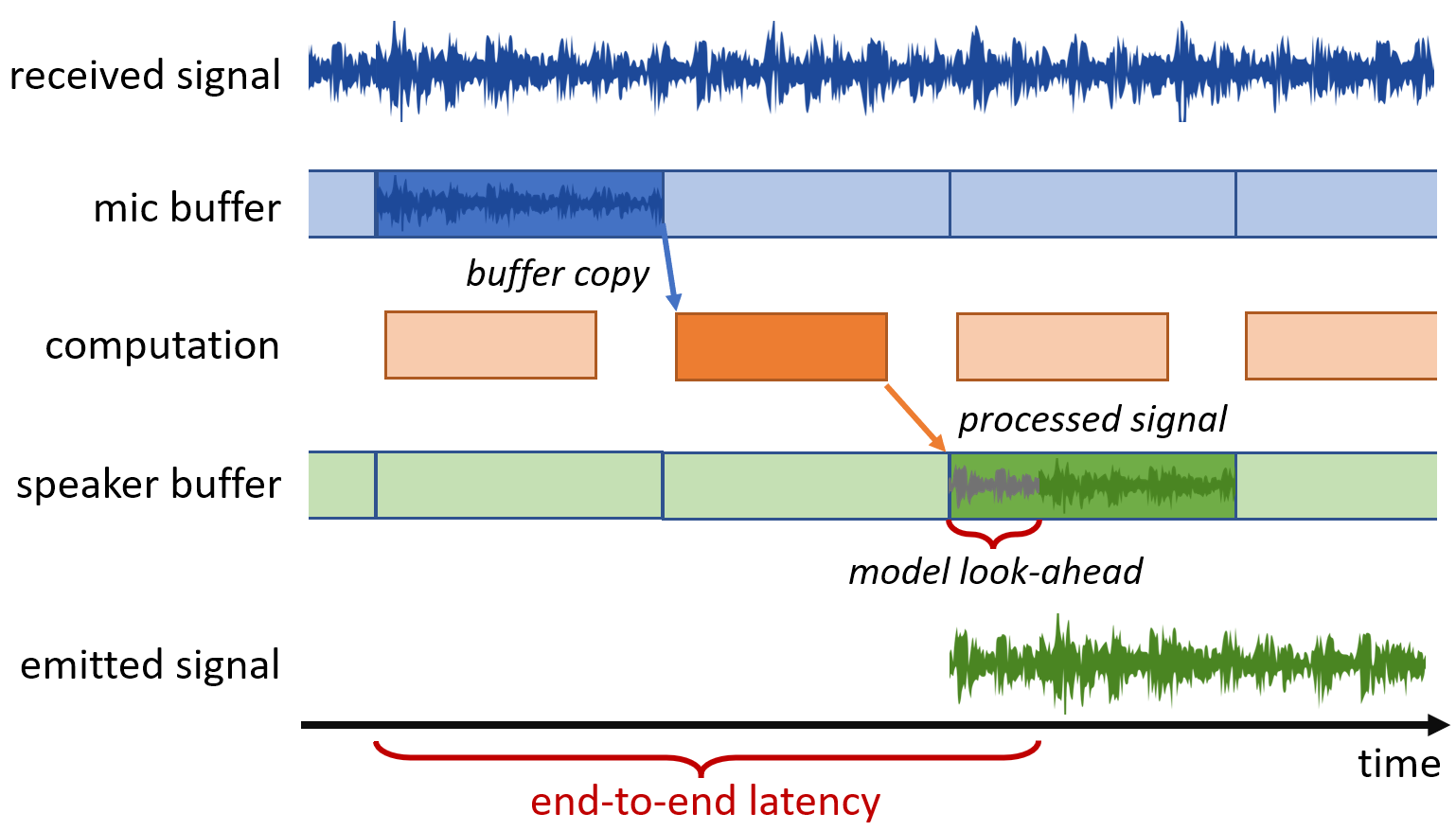}
    \vskip -0.1in
    \caption{\small{End-to-end latency for real-time hearing enhancement}}
    \label{fig:latency}
    \vskip -0.15in
\end{figure}

\begin{figure*}[t!]
    \centering
    \includegraphics[width=1\textwidth]{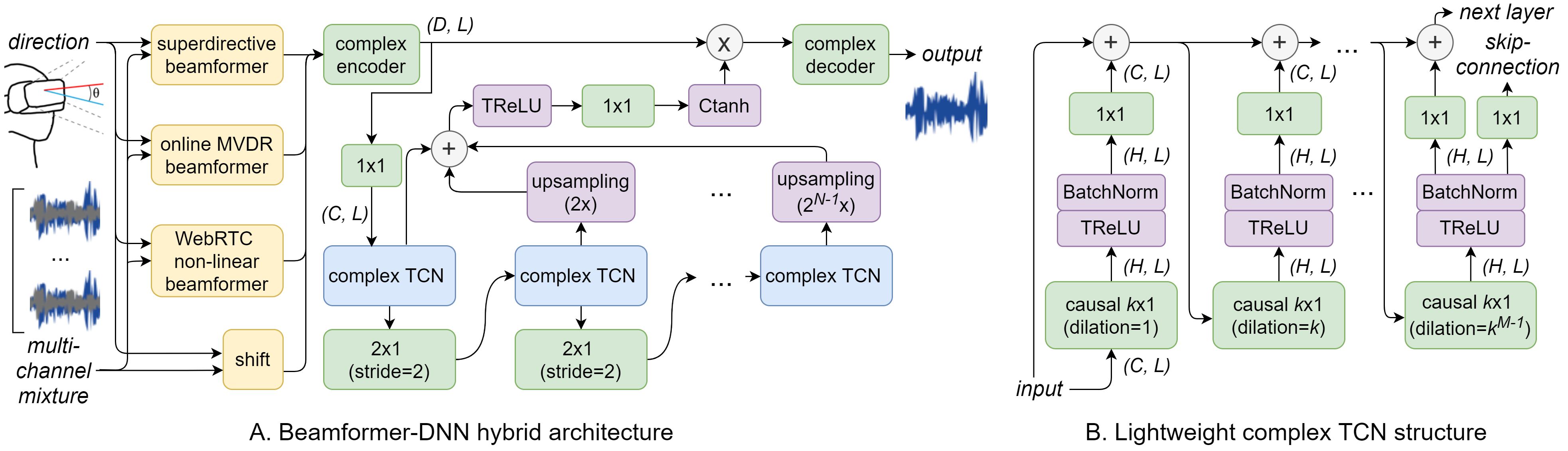}
    \vskip -0.1in
    \caption{The architecture of the HybridBeam system. (A): the end-to-end network diagram. (B): the structure of the simplified temporal convolutional network (TCN).}
    \label{fig:pipeline}
    \vskip -0.12in
\end{figure*}

\subsubsection{Improving MVDR with neural nets.}  \cite{zhang2021adl,xu2020neural,xiao2017time,tammen2019dnn} replace matrix inversion and PCA within a MVDR beamformer with a neural net. We take  the  inverse approach where we utilize the MVDR output for more efficient neural net feature extraction and design. It is also noteworthy that the structures of these prior designs are not causal in nature. Further, they are computationally expensive and can not run in real-time on a mobile CPU. Our joint beamformer-neural net approach instead reduces the complexity of the neural net using the features from the beamformers.}

\section{Method}

The problem of real-time direction hearing can be formulated as follows.  Say, we have $N$ sound sources $s_{1..N}$ emitted from  angles 
$\theta_{1..N}$ with respect to an  array with $c$ microphones. The signal received by the $i^{th}$ microphone is, 
$$y_i(t)=\sum_{j=1}^N \sum_{\tau=-\infty}^0 H_{i,j}(\tau)s_j(t-\tau)+N(t)$$ 
 Here, $N(t)$ is random noise and $H_{i,j}$ is the impulse response associated with sound source $j$ and microphone $i$ that captures  multi-path and reverberations.
At a given time $t$ and a known $\theta_k$, our goal is to  estimate the acoustic signal,  $s_k(t)$, emitted from the direction $\theta_k$, given $y(t-W)\cdots y(t+L)$, where $W$ is the reception field, and $L$ is a small look-ahead.

\subsection{Latency Requirements}

  Fig.~\ref{fig:latency} shows the composition of various sub-components that contribute to the  latency. First, the sound signals get sampled by the multiple  microphones and fed into a memory buffer. When the buffer is full, the data in the buffer is then processed by a computation program that consists of our neural network and signal processing techniques. The result, i.e., the speech from a specific direction, is then  transferred to be played back through the speakers.

Consequently, to reduce the end-to-end latency to less than 20~ms, we should 1) reduce the buffer size; 2) minimize the processing time; and 3) reduce the look-ahead duration for the model. Reducing all these parameters while achieving good performance is challenging for multiple reasons. First, a very short buffer size (say 3 ms) may cause  jitters due to the timing fluctuation in the operating system scheduling. A smaller buffer size also inversely increases with the frequency of the computation calls, each of which involves a constant overhead. Second, to ensure real-time operation, the computation block has to process the acoustic data from each buffer block within the duration of the block. That is, the computational time to process a 8 ms buffer should be less than 8~ms. Neural networks however are known for their heavy computational requirements, and none of existing  models are designed for such a small end-to-end latency and real-time signal processing on computationally constrained CPUs. Third, a large look-ahead can improve  source separation  but  also increases  latency.

\subsection{HybridBeam Architecture}

Fig.~\ref{fig:pipeline} shows the overall architecture of our hybrid model.
The input multi-channel signals from the microphone array and  the target angle $\theta_k$ are first passed to three lightweight beamformers that result in three different versions of the beamformed signals. These beamformed signals are concatenated with the original multi-channel signals and fed into our neural network that is designed to output the separated acoustic signal from the target direction $\theta_k$.

\subsubsection{Prebeamforming.} 
We use the following beamformers to extract features: a) superdirective beamformer~\cite{krim1996two} that is optimized under diffused noise; b) online adaptive MVDR beamformer~\cite{habets2010mvdr} that  extracts the spatial information from the past to suppress  noise and interference; and 3) WebRTC non-linear beamformer~\cite{kleijn2016methods} that enhances a simple delay-and-sum beamformer by suppressing the time-frequency components that are more likely noise or interference. These three statistical beamformers span the different classes of beamforming techniques from non-adaptive, adaptive and non-linear approaches. As a result, they provide a diversity of spatial information as input to the neural network. Moreover, they are all computationally efficient --- it takes 0.8ms to run all three beamformers for a 8~ms signal block on a mobile CPU.

Additionally, the input channels are shifted to aim at the input direction, so  that each channel samples the direct path of the signal at the same time in the  far-field: 
$$\hat{y}(f)=y(f)\exp(j2\pi f(t_i(\theta)-t_0(\theta)))$$
$t_i(\theta)$ is the time-of-arrival from direction $\theta$ on mic $i$.
These shifted channels along with the output of the  beamformers are concatenated together and fed into a neural network.

\subsubsection{Neural Network Model.} 
Our neural net is inspired by time-domain models like  Conv-TasNet~\cite{luo2019conv} and has a linear encoder, a linear decoder and a separator module, all with 1D convolutional layers. However, we make significant modifications to the network to reduce its memory footprint, memory copy overhead and to be computational lightweight that we describe in detail below.

\vskip 0.04in\noindent{\it Complex Tensor Representation.} We use complex tensors throughout our network to halve our model size ({{each parameter can be represented as $[R, -I; I, R]$, instead of full $2\times 2$ matrices.}}) while achieving a comparable accuracy. Complex representation is a powerful tool for acoustic signal processing. For example, complex multiplication capture rotation in the complex domain and can easily manipulate the signal phase. Thus, complex neural networks are found to be more effective for applications such as wireless communication~\cite{marseet2017application} and noise suppression~\cite{hu2020dccrn,bassey2021survey}. 
Compared to real-valued networks, complex representation also restricts the degree of freedom of the parameters by enforcing correlation between the real and imaginary parts, which enhances the generalization capacity of the model in other applications.
Besides the benefit of reduced model size, complex representation can be  especially important for beamforming since phase encodes essential spatial information.

Fully-complex neural networks however lack the capability to efficiently approximate \textit{conjugate} operation and \textit{phase scaling} where the phase of a complex number gets multiplied by a constant. To mitigate this, we insert an additional component-wise operation before each $\mathbb{C}\mathbf{ReLU}$ activation. We call them together a new $\mathbf{TReLU}$ activation. Specifically, we define  $\mathbf{TReLU}(\mathbf{x}_{c,t})$ as follows:
\begin{eqnarray*}
 ReLU(h_{c}^{(rr)}\Re(\mathbf{x}_{c,t})+h_{c}^{(ri)}\Im(\mathbf{x}_{c,t})+b_{c}^{(1)})\\+
 jReLU(h_{c}^{(ri)}\Re(\mathbf{x}_{c,t})+h_{c}^{(ii)}\Im(\mathbf{x}_{c,t})+b_{c}^{(2)})
\end{eqnarray*}
Here $\mathbf{x}$ is the complex input of the activation function, $c$, $t$ are the channel and time indices, respectively, and $h$, $b$ are parameters to train. Intuitively, the operation linearly transforms the 2D complex space that can simulate both conjugate and phase scaling, and then $ReLU$ activation is performed on real and imaginary parts independently. This is equivalent to the scaling operation in the complex batch normalization~\cite{trabelsi2017deep}, but we decouple it from the batch normalization which is moved after the $ReLU$ operation.
Note that the additional  computation of $\mathbf{TReLU}$ is negligible compared to the convolutional layers. 

\vskip 0.04in\noindent{\it Complex Masking.}  The separator outputs a complex mask range from 0 to 1 that is multiplied with the encoder output to feed into the decoder. {While the mask cannot go beyond 1, the trainable encoder and decoder could mitigate this limitation.} We apply a $tanh$ operation to the amplitude of the complex tensor while preserving the angle component:
$$\mathbf{Ctanh}(\mathbf{x})=tanh(||\mathbf{x}||)*\frac{\mathbf{x}}{||\mathbf{x}||}$$
\begin{figure}[t]
    \centering
    \includegraphics[width=0.37\textwidth]{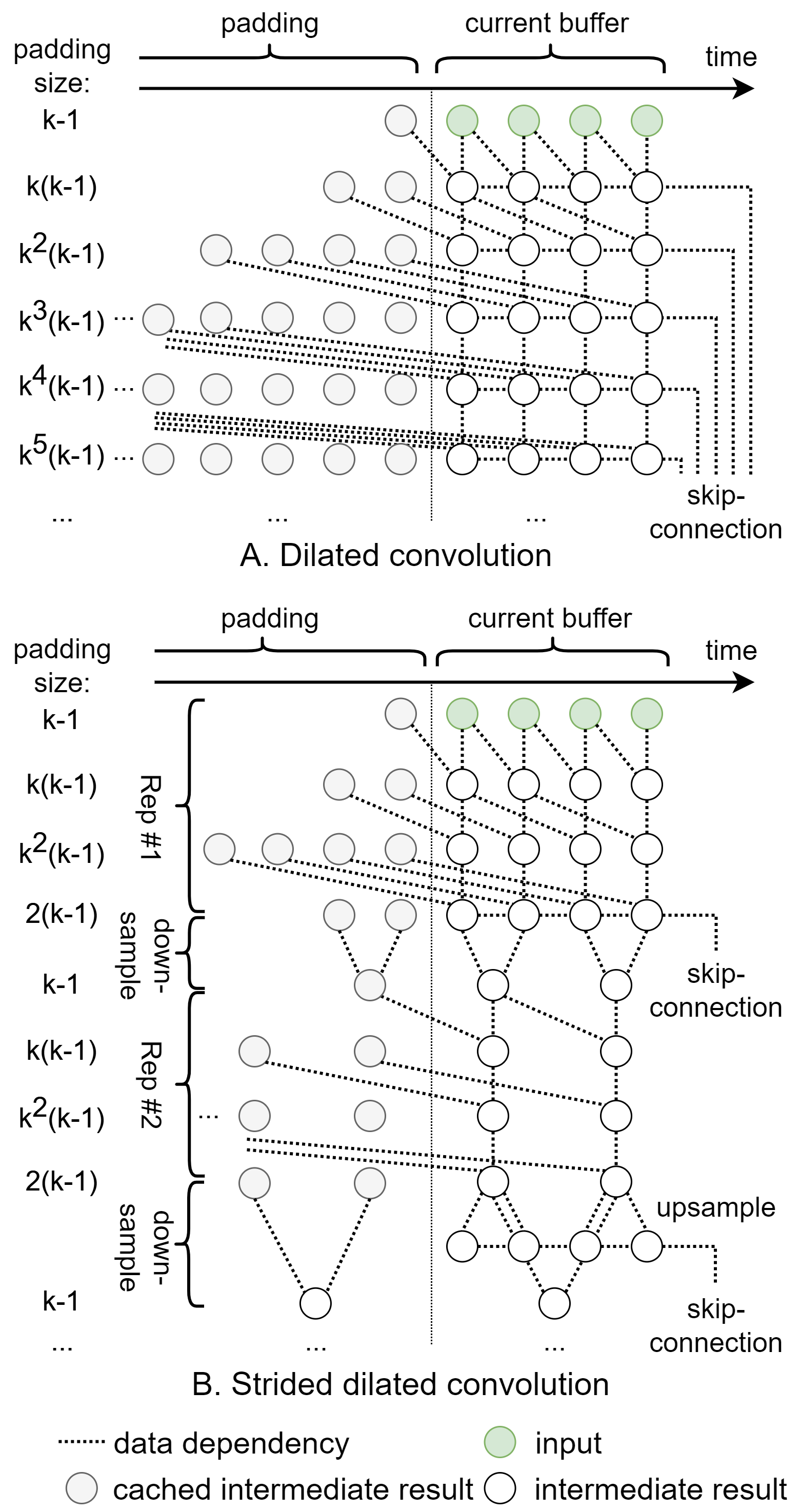}
    \vskip -0.1in
    \caption{The strided dilated convolution structure when $M=3$ and $k=2$. The total padding size is much reduced because of the downsampling layer, and skip-connections get upsampled accordingly before summed up.}
    \label{fig:wavenet}
    \vskip -0.18in
\end{figure}

\begin{table*}[t!]
\centering
\setlength{\tabcolsep}{4pt}
\begin{tabular}{|c|c|c|c|c|c|c|c|}
\hline
Configurations & Type & Hyperparameters  & recep. field & \# params & \# MAC/s & lookahead\\ \hline
HybridBeam  & hybrid        & k=4, N=3, M=3, H=64, C=64, D=256 & 0.22s & 0.72M     & 2.1G        & 1.5ms        \\
HybridBeam+ & hybrid         & k=3, N=4, M=4, H=96, C=64, D=256  & 0.61s & 1.1M    & 2.8G          & 1.5ms      \\ \hline
TSNF & DNN & N=512, L=16, H=512, Sc=128, X=8, R=3  & 0.77s  & 5.2M      & 10.4G        & 390ms       \\
TAC-F & DNN & H=32, L=64, W=64,K=64 & $\infty$ & 2.8M & 14.5G    & $\infty$ \\ \hline
Online MVDR & traditional   & -    & -         & -   &  -     & 0           \\
Mod. TSNF & DNN    & N=512, L=32, H=512, Sc=128, X=8, R=3  & 0.77s  & 5.2M      & 10.4G        & 1.5ms       \\
Mod. TAC-F & DNN    & H=32, L=64, W=64,K=64  & $\infty$  & 2.3M      & 11.6G         & 4ms        \\ \hline
\end{tabular}
\vskip -0.1in
\caption{The specification of our model and baselines}
\label{tab:spec}
\vskip -0.15in
\end{table*}
\noindent{\it Strided Dilated Convolution.}
Temporal convolutional networks (TCN) utilize causal dilated convolution to efficiently enlarge the audio receptive field and achieve good separation performance~\cite{luo2019conv}. For real-time applications, while intermediate convolution results from the past can be cached for fast computation, the memory copy overhead is significant when our latency requirement is in the order of milliseconds.  For example, Fig.~\ref{fig:wavenet}A shows how the Conv-TasNet~\cite{luo2019conv} architecture processes a stream of audio buffers. The input padding for each convolution layer contains temporal information that is computed while processing the previous buffers, and the  input is shifted left and set as the new padding for the new buffer input. The shifted padding increases exponentially  for latter layers due to a large receptive field. The shift operation is usually implemented using array copy and the state-of-the-art Conv-TasNet model  requires 25~MB memory copy per input, that takes approximately 10~ms on a Raspberry Pi.

To reduce the memory copy overhead, we  design a combination of dilated and strided convolution stacks. As shown in Fig.~\ref{fig:wavenet}B, the network consists of a stack of $N$ TCNs. Between each TCN which contains $M$ dilated convolution layers, we add a 2x1 convolution layer with $stride=2$ to downsample the signal and effectively reduce the size of the padding for the following layers. The skip-connections get upsampled using the nearest neighborhood method accordingly to the original sampling rate before summed up. {Similar concept has been explored in \cite{tzinis2020sudo}, and we extend it to the TCN structure.} Compared to the original TCN stacks in \cite{luo2019conv}, our strided dilated convolution technique requires the same $O(k^MN)$ padding to achieve a much larger $O(k^M2^N)$ receptive field instead of $O(k^MN)$. For the  specific parameters we implement, this reduces memory copy by $>$90\%.

\vskip 0.04in\noindent{\it Simplified Temporal Convolutional Network.} We further simplify the original TCN design. First, we use conventional convolution instead of depthwise separable convolution (D-conv) which has lower MACs for the same channel capacity but is usually memory-bounded and 4-8x less efficient than normal convolution operations on mobile CPUs~\cite{zhang2020high}. In our experiments with a few state-of-the-art mobile DNN inference engines, normal convolution was much  faster than D-conv.
Second, we apply skip-connections on only the last convolution layer for each TCN stack, as shown in Fig.~\ref{fig:pipeline}B. We found that this reduction of skip-connections  reduces computation by approximately 20\%. 
Third, we relax the dilation growth factor $k$ to more than 2. Since our receptive field is $O(k^M2^N)$, by increasing $k$, we would need less $M$ and $N$, and so lower number of layers, to achieve the same receptive field.

\section{Evaluation}
We prototype and train the neural network in PyTorch~\cite{paszke2019pytorch}, and rewrite the model in TensorFlow~\cite{abadi2016tensorflow} as TensorFlow supports NHWC tensor layout which is faster on mobile CPUs. The model get converted to the input formats of two DNN inference engines, the MNN from Alibaba and Arm NN. The latter supports NEON and 16~bit float (FP16) primitives for ARMv8.2 CPUs. 
We use PulseAudio to access the microphones in real time  and use a sampling rate of 16~kHz and 16~bit bitwidth.

\begin{table}[t]
\centering
\begin{tabular}{|c|ccc|}
\hline
Configuration & SI-SDRi & SDRi & PESQ \\ \hline
HybridBeam         &  11.3~dB & 10.1~dB & 1.61       \\
HybridBeam+         & \textbf{13.3~dB}  & 11.4~dB & 1.92      \\ \hline
HybridBeam+ w/ FP16 & 13.1~dB & 11.2~dB  & 1.92\\
HybridBeam+ w/o BF         & 10.9~dB    & 9.7~dB & 1.70       \\ \hline
Online MVDR   & 5.2~dB        & 5.9~dB & 1.52    \\
Mod. TSNF    & 13.1~dB        & \textbf{11.6~dB} & \textbf{1.94}      \\
Mod. TAC-F    & 12.1~dB        & 11.1~dB & 1.77  \\ \hline
\end{tabular}
\vskip -0.1in
\caption{
\label{tab:synthetic_result}The quantitative performances of our method compared with baselines using a circular 6-mic array}
\vskip -0.15in
\end{table}

\subsection{Simulated Dataset}
\begin{figure}[t]
    \centering
    \includegraphics[width=0.42\textwidth]{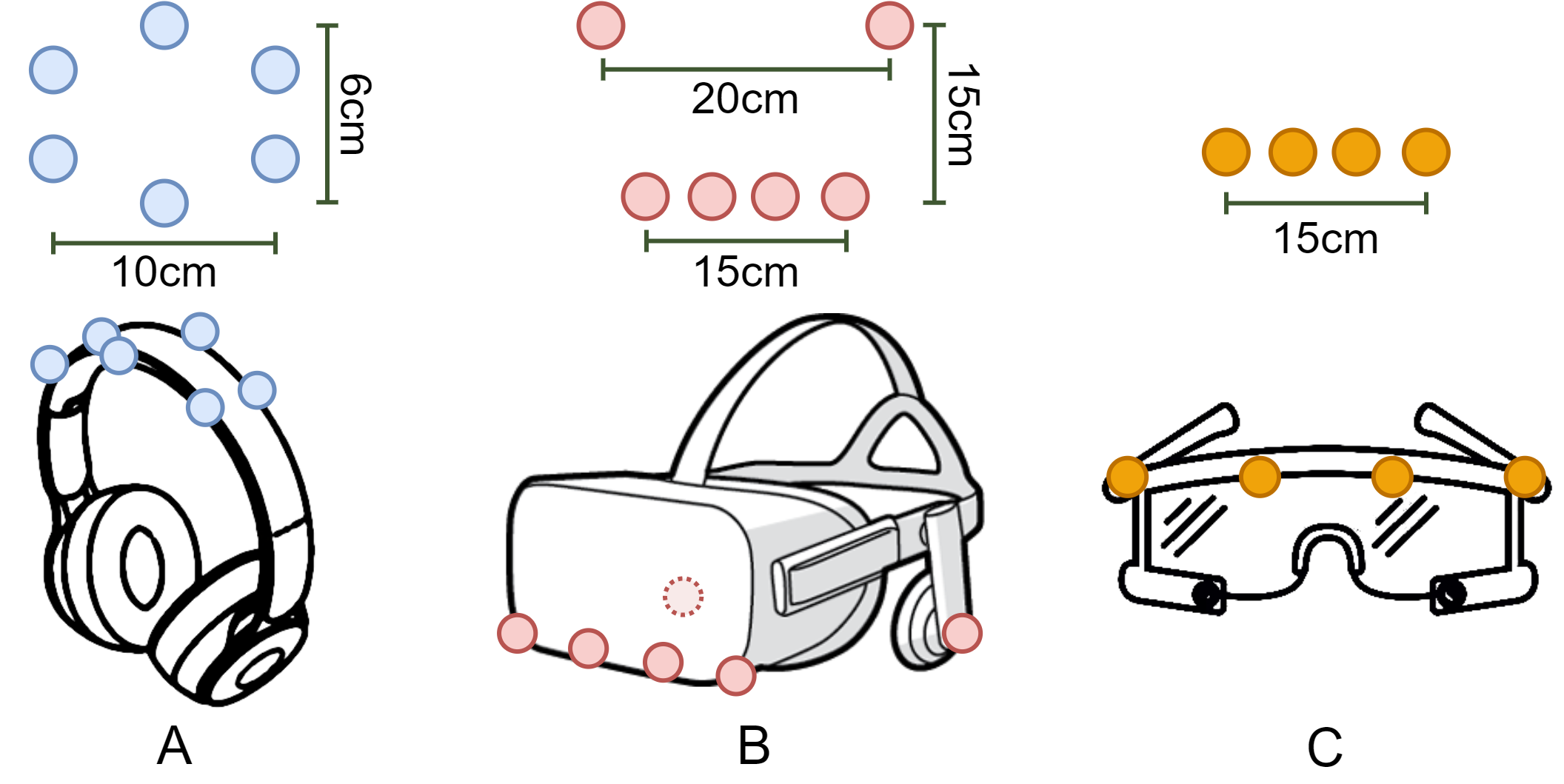}
    \vskip -0.1in
    \caption{Potential mic-array layouts. (A): six-mic hexagon array on top of a headphone; (B): five-mic sub-array of a six-mic array (the microphone on the left/right ear is disabled when the input direction is on the opposite  side) on an AR headset; (C) four-mic linear array on a pair of smart glasses.}
    \vskip -0.1in
    \label{fig:miclayout}
\end{figure}

To gather a large amount of training data, we use software to simulate random reverberate noisy rooms using the image source model~\cite{scheibler2018pyroomacoustics}. The rooms are simulated using absorption rates of real materials and a maximum RT60 of 500~ms. By default, we use a virtual 6-mic circular array with a radius of 5~cm. The distance between the virtual speakers and the microphone array is at least 0.8~m, and the direction of arrival differences of the speakers are at least $10^{\circ}$. The input direction is modeled as the groundtruth plus a random error less than $5^\circ$ simulating the gaze tracking measurement error \cite{sipatchin2020accuracy}.  We place virtual speakers at random locations within the room playing random speech utterances from the VCTK dataset~\cite{veaux2017cstr}, meanwhile simulating diffused noise from the MS-SNSD dataset~\cite{reddy2019scalable} and WHAM! dataset~\cite{wichern2019wham}. The combined speech power to noise ratio is randomized between $[5, 25]$~dB. 10\%, 40\%, 40\%, 10\% of the generated clips consists of 1-4 speakers, respectively, and we apply random gain within $[-5, 0]$~dB to each speaker. We guarantee that there exists speech utterance overlap for 2-4 speaker scenarios. We render the synthetic audio and generate 4s clips. We generate a total of 8000 clips as training set, 400 clips as validation set, and 200 clips as test set. No speech clips or noise has appeared in more than one of these three sets. To evaluate the performance on different microphone number and array layouts on various wearable form factors, we additionally generate datasets using three custom microphone array layouts on a virtual reality (VR) headset as shown in Fig.~\ref{fig:miclayout}.

\begin{table}[t]
\centering
\setlength{\tabcolsep}{4pt}
\begin{tabular}{|c|cccc|c|}
\hline
\multicolumn{1}{|c|}{\multirow{2}{*}{\#mic}} & \multicolumn{4}{c|}{\# sources} &\multicolumn{1}{|c|}{} \\
\multicolumn{1}{|c|}{}                        & 1      & 2      & 3     & 4 &  \multicolumn{1}{|c|}{Overall}   \\ \hline
4                                             & 11.9~dB       & 10.8~dB        & 10.5~dB      & 8.4~dB & 10.6~dB     \\
5                                             & 11.8~dB       & 12.0~dB        & 11.9~dB      & 9.1~dB & 11.7~dB     \\
6                                             & 11.6~dB       & 13.4~dB        & 13.5~dB      & 10.2~dB & 12.9~dB     \\ \hline
\end{tabular}
\vskip -0.1in
\caption{
\label{tab:micsource}SI-SDRi performance using three custom microphone array layout under different number of sound sources}
\vskip -0.1in
\end{table}

\subsection{Model Specification}
We use two specifications for our model. The encoder and decoder  both have a kernel size of 32 and a stride of 8. The rest of the hyperparameters are listed in Table~\ref{tab:spec}. The lookahead comes from the transposed convolution in the decoder. We use three baselines for reference: 1) traditional, online MVDR beamformer; 2) modified Temperal Spatial Neural Filter (TSNF)~\cite{gu2020temporal}, where we replace the TasNet structure with a causal Conv-TasNet structure and use the identical encoder as ours to achieve the same lookahead duration; and 3) modified TAC-FasNet (TAC-F)~\cite{luo2019fasnet}, where we replace bidirectional RNN with directional RNN for causal construction, conduct the same alignment operation to the multi-channel input before feeding into the network, and output only one channel.

\begin{figure}[t]
    \centering
    \includegraphics[width=0.48\textwidth]{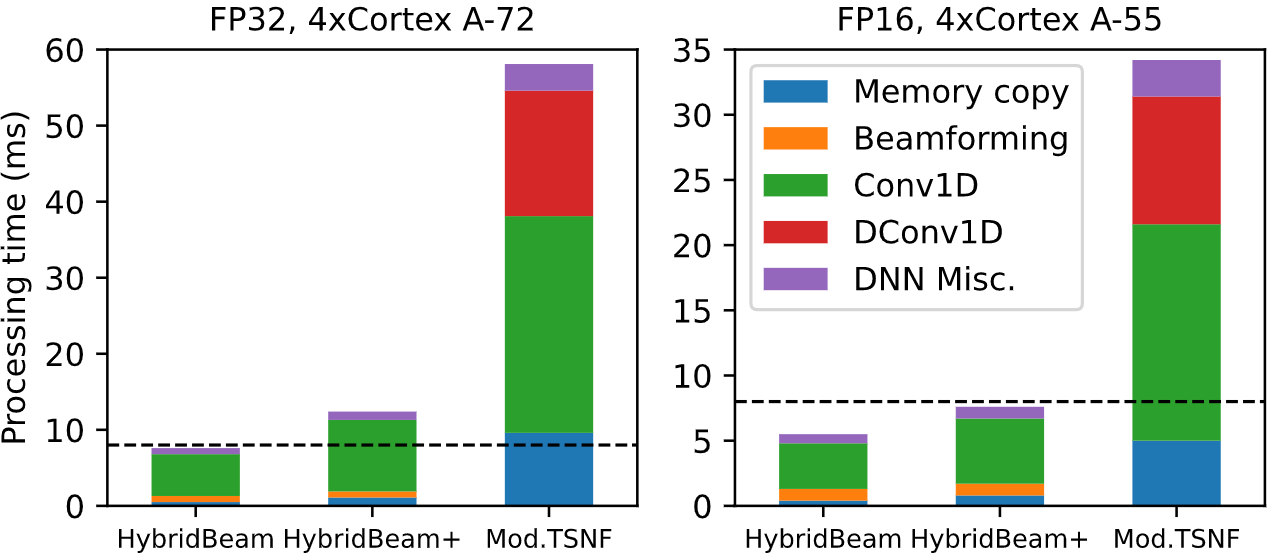}
    \vskip -0.1in
    \caption{The processing time composition. The dashed line is   maximum processing time to achieve real-time operation.}
    \label{fig:latencycmp}
    \vskip -0.1in
\end{figure}

\subsection{Training Procedure}
 When synthesizing each training audio clips, we additionally synthesize another version where only one of the sound source and the first microphone are present, and no reverberation is rendered. This version is used as the groundtruth  when the direction input is the direction of the present source. Hence, our model is trained to {\it simultaneously} do de-reverberation, source separation and noise suppression. 
We use a 1:10 linear combination of scale-invariant signal-to-distortion ratio (SI-SDR)~\cite{le2019sdr} and mean L1 loss as training subjective, where the former is used to measure the speech quality, and the latter regulates the output power to be similar to the groundtruth.

\begin{table}[t]
\centering
\begin{tabular}{|c|cccc|}
\hline
RT60(s) & $<0.2$ & $0.2-0.4$ & $0.4-0.6$ & $>0.6$ \\ \hline
SI-SDRi & 13.5~dB & 13.3~dB & 12.8~dB & 11.1~dB \\ \hline
\end{tabular}
\vskip -0.1in
\caption{
\label{tab:reverb}SI-SDRi under different reverberation time}
\vskip -0.1in
\end{table}

\begin{table}[t]
\centering
\setlength{\tabcolsep}{4pt}
\begin{tabular}{|c|cccc|}
\hline
Removed BF & None & SD & MVDR & Nonlinear \\ \hline
SI-SDRi & 13.3~dB & 13.1~dB & 12.0~dB & 12.4~dB \\ \hline
\end{tabular}
\vskip -0.1in
\caption{
\label{tab:ablation}SI-SDRi when we remove a beamformer.}
\vskip -0.1in
\end{table}

\begin{table}[t]
\centering
\setlength{\tabcolsep}{4pt}
\begin{tabular}{|c|ccccc|}
\hline
diff (°) & 10-20&20-30&30-40&40-50&$>$50\\
\hline
w/ error (dB) & 7.4 & 10.6&13.3&13.5&13.7\\
\hline
 w/o  error (dB)&7.9& 11.7& 13.5& 13.8 & 13.9
 \\ \hline
\end{tabular}
\vskip -0.1in
\caption{
\label{tab:anglediff}Different direction differences between 2 sources.} 
\vskip -0.2in
\end{table}

\subsection{Network Evaluation on Synthetic Datasets}

Table~\ref{tab:synthetic_result} shows the  SI-SDR  (SI-SDRi) and SDR improvements (SDRi). We find that all DNN-based approaches outperform traditional MVDR beamformer. Our HybridBeam+ model that uses a slightly larger model achieves comparable results with the causal and low lookahead version of  \cite{gu2020temporal}, but using significant less number of parameters and computation. 
We additionally evaluate two variants of our large model. First, we use 16 bit float format (FP16) instead of 32 bits and see only a 0.2~dB drop in both SI-SDRi and SDRi. Using FP16 drastically reduce the inference time on platforms that support native FP16 instructions. 
Second, we remove the three beamformers  and retrain the network. The SI-SDRi  drops by more than 2~dB, which shows the usefulness of prebeamforming.
 {We used bootstrapping sampling techniques to evaluate our testset 100 times. The 25th, 50th and 75th percentile are 12.99 dB, 13.32 dB and 13.61 dB respectively on our HybridBeam+ model.}

Table~\ref{tab:micsource} shows the SI-SDRi results on the custom microphone array layouts under 1-4 sources. Note that the training set contains all 1-4 source scenarios and each row of the table shows the performance on the testset under a same trained model.
We see that adding microphones consistently improves the result of more than one source scenarios. 

We also evaluate the performance with  different reverberation time (RT60) in Table.~\ref{tab:reverb}. We see a performance degradation with a RT60 greater than 0.6~s, likely due to a limited receptive field.  Table~\ref{tab:ablation} shows an ablation study, where we remove one out of the three beamformers and retrain the network. {We also evaluated HybridBeam+ with only one reference channel (the first microphone channel) without shifting along with the output
of the beamformers as input. The resulting SI-SDRi is only
0.2~dB lower, which indicates the usefulness of pre-beamforming.} Table~\ref{tab:anglediff} shows how our model performs with different directional differences between two sources. The separation performance increases as the angular difference between the sources increases. When there is no direction error in the input, the SI-SDRi improves for smaller angular differences.

We also compared our results with two real-valued networks with the same HybridBeam+  structure: (1) a real-valued version  trained with dimensions adjusted to match the number of trainable parameters in our complex-valued network and 2) a real-valued network  constructed with a same number of CNN channels (thus, twice the number of trainable parameters). The first network had a 0.5~dB SI-SDRi drop compared to our complex network. The second topline network achieved a 0.6~dB SI-SDRi gain. This shows that  our complex-valued network has a good tradeoff.

\subsection{On-device Latency Analysis}

We deploy the models on two mobile development boards to measure the processing latency: a Raspberry Pi 4B with a four-core Cortex A-72 CPU and a four-core low-power Cortex A-55 developement board which support FP16 operations, both running at 2~GHz. The former is a popular \$35 single-board computer, and the latter CPU is designed for low-power wearable devices and efficient cores on smartphones for lightweight tasks like {checking emails}.

We run the model in real-time and the buffer size is set to 128 samples (8~ms). Recall from Fig.~\ref{fig:latency} that the processing time should be less than 8~ms to guarantee real-time operation.  
Fig.~\ref{fig:latencycmp} shows the processing time of  of our models as well as the causal version of \cite{gu2020temporal}. We find that with a comparable source separation performance, inference using our model takes much less time. Specifically, memory copy overhead is significantly reduced because of the strided dilated convolution, so does  computation because of an overall smaller model with vanilla convolution. Finally, with a  lookahead of 1.5~ms, our  models can run on our two platforms  in real-time with a 17.5~ms end-to-end latency. 

\subsection{Network Evaluation on Hardware Data}

\subsubsection{Hardware dataset.} 
\begin{figure}[t]
    \centering
    \includegraphics[width=0.23\textwidth]{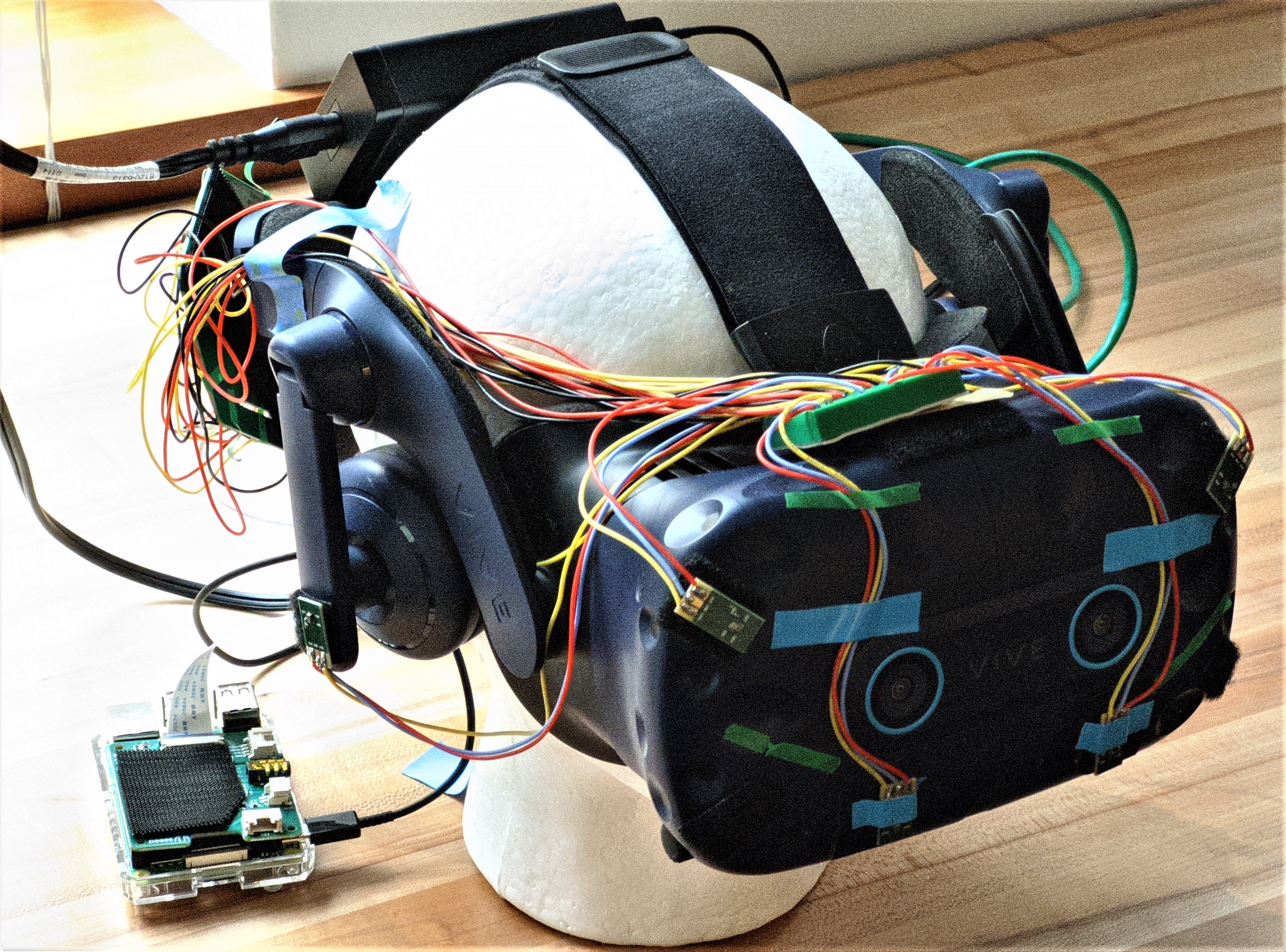}
    \vskip -0.1in
    \caption{Gaze-controlled directional hearing AR prototype.}
    \label{fig:dummyhead}
\end{figure}

To evaluate model generalization, we  implement a headset prototype and test with actual hardware. We modify a Seeed ReSpeaker 6-Mic Circular Array kit and  place the microphones in the configuration in  Fig.~\ref{fig:miclayout}B around a HTC Vive Pro Eye VR headset in   Fig.~\ref{fig:dummyhead}. The headset's gaze tracker provides the direction of arrival for our model. In addition to generating synthesized data using the above procedure but with the actual microphone layout, we also collect hardware data in two  different rooms: {one large, empty and reverberate conference room (approximately $5\times 7m^2$), denoted as \textit{Room La}, and one smaller, regular room with desks (approximately $3\times 5m^2$), denoted as \textit{Room Sm}.} The playback speech is from the same VCTK dataset but is played from a portable Sony SBS-XB20 speaker. We place the speaker at 1~m and different angles within $-75^{\circ}$ to $75^{\circ}$. 
We calibrate the speaker-microphone delay and phase distortions in an anechoic chamber using a chirp signal and apply the calibration to the original  signal. 
After data collection, we randomly add two recordings whose direction of arrival difference is more than $10^{\circ}$ as the mixture signal. We pick the calibrated original speech and the direction of arrival of one of them as groundtruth and input direction to our model.

We use our model to test on hardware datasets collected in the conference rooms. We first see if training on only synthesized data can generalize to hardware data. {We choose the best baseline on the synthetic datasets for comparison.} Table~\ref{tab:realdataset} shows that the previous work generalizes poorly and does not work in real data. Manual inspection indicates that the the model sometimes predicts wrong sound sources. This is mostly because the features used by TSNF is highly affected by noise and interference and is not robust in real-world scenarios. In contrast, our model generalizes and outperforms MVDR baseline. Our hypothesis is that our model focuses on improving the already beamformed signals instead of deciding which source to separate, which is a harder problem. 
We also mix the 50\% actual recordings with 50\% synthesized data as the training set and test on the recordings in another room. The model  performs better and achieves another 3~dB gain, {regardless of the room acoustic properties.}

\begin{table}[t]
    \centering
    \setlength{\tabcolsep}{3pt}
    \begin{tabular}{|c|c|c|c|}
    \hline Model & Training & Test & SI-SDRi \\ \hline
        \multirow{2}{*}{MVDR} & \multirow{2}{*}{None} & Room La & 4.5~dB \\
        & & Room Sm & 4.2~dB \\ \hline
        TSNF & \multirow{2}{*}{Synthetic} & \multirow{2}{*}{Room La} & 3.1~dB \\
        Mod. TSNF & & & -1.6~dB \\ \hline
        \multirow{3}{*}{HybridBeam+} & Synthetic & Room La & 5.6~dB \\
         & Synthetic+Room Sm & Room La & 8.7~dB \\ 
         & Synthetic+Room La & Room Sm & 8.2~dB \\ \hline
    \end{tabular}
    \vskip -0.1in
    \caption{Performance using real world data.}
    \vskip -0.2in
    \label{tab:realdataset}
\end{table}
\section{Conclusion and Limitations}
We believe that while this paper makes  important contributions in demonstrating low-complexity neural networks for directional hearing, there is scope for improvements.

\textbf{Larger real-world datasets.} Our model can generalize to real recordings using synthetic training data, but is not as good as using synthetic test set, because of audio distortion like hardware nonlinearity and audio refraction. Generalization could be further improved with a larger scale real-world  data collection under various rooms and devices.

\textbf{Binaural settings.} When we  retrain the network with 2 microphones and up to 2 sources, we had an overall SI-SDRi performance of only 6.1~dB. This is likely because we depend on the beamformers to reduce the computational complexity of the neural networks.  With 2 microphones, superdirective, MVDR and WebRTC beamformers only provided  1.8, 1.9, -0.6~dB respectively.  Binaural beamformers~\cite{liao2015effective,srinivasan2008low,hadad2017comparison} can potentially be used to improve performance.

\textbf{Specialized hardware.} We use low-power CPU to deploy our model because of its extensive adoption in wearables and high flexibility. Recent low-power DNN accelerators can run DNN more efficiently~\cite{wang2019benchmarking} and can be used to further improve our DNN efficiency, while running the beamformers on a CPU or  microcontroller.

\textbf{Lower latency.} While our target end-to-end latency can be useful for  wearables, close-canal hearing aids or hearing devices with active noise cancellation, an open-canal setting requires more stringent latency requirements~\cite{stone2008tolerable}. This may need hardware-software co-design.

{\textbf{Runtime optimizations.}  Techniques like network pruning and quantization  may  potentially help  existing models  run on-device. Those techniques however may also help our model  run on even more resource-constraint hardware.}

\section{Acknowledgments}
We thank Les Atlas, Steve Seitz, Laura Trutoiu and Ludwig Schmidt for their important  feedback on this work.

\bibliography{reference}

\end{document}